\documentclass[letterpaper,twocolumn,english,prl,floatfix,superscriptaddress,showpacs,preprintnumbers]{revtex4}
\usepackage[T1]{fontenc}
\usepackage[latin9]{inputenc}
\usepackage{babel}
\usepackage{float}
\usepackage{amsthm}
\usepackage{amsmath}
\usepackage{amssymb}
\usepackage{graphicx,epsf}
\usepackage[unicode=true, pdfusetitle,
 bookmarks=true,bookmarksnumbered=false,bookmarksopen=false,
 breaklinks=false,pdfborder={0 0 1},backref=false,colorlinks=false]
 {hyperref}

\makeatletter

\floatstyle{ruled}
\newfloat{algorithm}{tbp}{loa}
\floatname{algorithm}{Algorithm}

\@ifundefined{textcolor}{}
{%
 \definecolor{BLACK}{gray}{0}
 \definecolor{WHITE}{gray}{1}
 \definecolor{RED}{rgb}{1,0,0}
 \definecolor{GREEN}{rgb}{0,1,0}
 \definecolor{BLUE}{rgb}{0,0,1}
 \definecolor{CYAN}{cmyk}{1,0,0,0}
 \definecolor{MAGENTA}{cmyk}{0,1,0,0}
 \definecolor{YELLOW}{cmyk}{0,0,1,0}
 }
\theoremstyle{plain}
\theoremstyle{plain}

  \theoremstyle{plain}

\usepackage{bbm}
\def\id{\mathbbm{1}}

\makeatother

\newcommand\beq{\begin{equation}}
\newcommand\eeq{\end{equation}}
\newcommand\EE{{\cal E}}
\newcommand\al{\alpha}
\newcommand\la{\lambda}
\newcommand\om{{\omega}}
\newcommand\si{{\sigma}}
\newcommand\bra[1]{\left\langle#1\right|}
\newcommand\ket[1]{\left|#1\right\rangle}
\newcommand\vev[1]{{\left\langle{#1}\right\rangle}}
\newcommand\Tr{{\rm Tr}}

\renewcommand{\SS}{\mathcal{S}}
\begin{document}
\preprint{UdeM-GPP-TH-09-182}

\title{Decoherence suppression via environment preparation}

\author{Olivier Landon-Cardinal}
\email{olivier.landon-cardinal@umontreal.ca}
\affiliation{D\'epartement de physique, Universit\'e de Montr\'eal, C.P. 6128, Succ.~Centreville, Montr\'eal, QC H3C 3J7 Canada}
\affiliation{D\'epartement IRO, Universit\'e de Montr\'eal, C.P. 6128, Succ.~Centreville, Montr\'eal, QC H3C 3J7 Canada}

\author{Richard MacKenzie}
\email{richard.mackenzie@umontreal.ca}
\affiliation{D\'epartement de physique, Universit\'e de Montr\'eal, C.P. 6128, Succ.~Centreville, Montr\'eal, QC H3C 3J7 Canada}

\date{\today}

\begin{abstract}

To protect a quantum system from decoherence due to interaction with its environment, we investigate the existence of initial states of the environment allowing for decoherence-free evolution of the system. For models in which a two-state system interacts with a dynamical environment, we prove that such states exist if \emph{and only if} the interaction and self-evolution Hamiltonians share an eigenstate. If decoherence by state preparation is not possible, we show that initial states minimizing decoherence result from a delicate compromise between the environment and interaction dynamics.




\end{abstract}


\maketitle

{\em Introduction~~~}
Decoherence provides an elegant framework to explain why an open quantum system
coupled to its environment will exhibit a set of preferred states,
usually ruling out a coherent superposition of arbitrary states. In
 \cite{Einselection}, Zurek showed that even if the superposition
principle treats all quantum states equally, the interaction between the quantum system
and its environment would select a restricted number
of {}``pointer'' states (einselection) and destroy the phase coherence 
of superpositions of those pointer states (decoherence). This phenomenon presents a formidable challenge for such applications as quantum computation.
In this simplest model of decoherence, it was readily realized that initial
states of the environment exist that allow for decoherence-free unitary evolution
of the quantum system. These peculiar states were usually neglected
on the basis that {}``in realistic cases, such highly ordered initial
states [$\dots$] are unlikely to be relevant'' \cite{Schlosshauer2007}
or because the environment self-evolution would preclude such a unitary evolution \cite{LesHouchesPazZurek}.

In this paper, we investigate the conditions under which such
initial states of the environment do exist in a framework where the quantum
system interacts with its environment and the environment also evolves
by itself. The results obtained underline the crucial role of the
environment's self-evolution. The ability to identify and prepare such special
initial states could be used in order to store quantum
states. Indeed, even if the environment dynamics cannot be controlled, it might
be possible to prepare it in a specific initial state. However, our
results restrict what can be expected from such a technique. 

More precisely, we obtain a mathematical condition
for the existence of an initial state allowing decoherence-free evolution
in the presence of an interaction Hamiltonian and a self-evolution of
the environment, stated in terms of the structure of
the two Hamiltonians. Next, we analyze in detail
a particular model, which is an extension of the model introduced by Zurek in \cite{Einselection}. Finally, we assess the impact of imperfect state preparation and discuss how to choose an initial state that minimizes decoherence when state preparation cannot avoid it altogether.

{\em The model~~~}
We consider a quantum system $\SS$ and its environment $\EE$
with dynamics described by a Hamiltonian of the form
\beq
\label{hamiltonian}
H=S\otimes\tilde{H}+\id\otimes H_{\EE}.
\eeq
The Hermitian operator $S$ acts in the Hilbert space of $\SS$, which we take to be two-dimensional. 
The quantum system is thus taken to be a quantum bit (qubit) \cite{NielsenChuang}.
the Hermitian operators $\tilde H$ and $H_{\EE}$ act in the Hilbert space of the environment.

The total Hamiltonian \eqref{hamiltonian} induces \emph{pure dephasing} and is typical of a coupling between the system and the environment that commutes with the self-evolution of the system \cite{Hornberger2009}.
 
Without loss of generality, we can assume $S=\si^z$. 
Its eigenstates, of eigenvalues $\pm1$, are written $\ket{0}$ and $\ket{1}$, respectively.

Suppose that the global system is in a product state at $t=0$:
\[
\ket{\Psi(0)}=\ket{\psi(0)}\otimes\ket{I}
\]
where $\ket{\psi(0)}=a\ket{0}+b\ket{1}$ is an arbitrary normalized pure
qubit state. At a later time $t$, the state evolves to
\[
\ket{\Psi(t)}=a\ket{0}\otimes\ket{\varepsilon_{0}(t)}+b\ket{1}\otimes\ket{\varepsilon_{1}(t)}
\]
where
\beq
\label{schr1}
i\frac{\mbox{d}}{\mbox{dt}}\ket{\varepsilon_{k}(t)}=H_{k}\ket{\varepsilon_{k}(t)},\qquad k=0,1,
\eeq
and  where we have defined $H_0\equiv H_{\EE} + \tilde{H}$ and $H_1\equiv H_{\EE} - \tilde{H}$
 with initial condition $\ket{\varepsilon_{0}(0)}=\ket{\varepsilon_{1}(0)}=\ket{I}$.

Since $\ket{\Psi(t)}$ is no longer a product state in general, the reduced density matrix $\rho(t)$ of the quantum system $\SS$ no longer describes a pure state: the system has decohered. To quantify this, the off-diagonal elements of $\rho(t)$ in
the basis $\{\ket{0},\ket{1}\}$ are reduced by a factor 
\[
r(t)=\vev{\varepsilon_{0}(t)|\varepsilon_{1}(t)}\mbox{.}
\]
Essentially, $r(t)$ quantifies how distinguishable
the states $\ket{\varepsilon_{0}(t)}$ and $\ket{\varepsilon_{1}(t)}$
are. If they differ only by a phase, i.e. $|r(t)|=1$, they are indistinguishable and the system $\SS$ remains in (or has returned to) a pure state. If $r(t)=0$, they can be distinguished with certainty, decoherence is complete, and the
state of the quantum system reduces to a statistical mixture $|a|^2\ket{0}\bra{0}+|b|^2\ket{1}\bra{1}$.
In general, the system has only undergone partial decoherence and a straightforward calculation
shows that its purity is
\[
\mathcal{P}(t)=\Tr\rho^{2}(t)=1-2|a|^2|b|^2(1-|r(t)|^{2})
\]
where $|r(t)|^{2}$ is the Loschmidt echo \cite{Cucchietti2005}.

Thus, a qubit prepared in an arbitrary coherent superposition of pointer
states will remain pure if and only if the environment can be prepared in a state $\ket{I}$ for which
\begin{equation}
|r(t)|=1\mbox{ for all }t
\label{eq:coherence_criterion}
\end{equation}
i.e., for which $\ket{\varepsilon_{0}(t)}$ and $\ket{\varepsilon_{1}(t)}$
are the same state up to a time-dependent phase at all time. 


{\em Condition for coherent evolution~~~}
Given that the state evolves according to \eqref{schr1}, what can be said of the Hamiltonians $\tilde{H}$ and $H_{\EE}$ in order that for some initial state $\ket{I}$, $\ket{\varepsilon_{0}(t)}$ and $\ket{\varepsilon_{1}(t)}$ satisfy the the coherence criterion \eqref{eq:coherence_criterion}? 
The factor $r(t)$ is given by
\[
r(t)=\bra{I} e^{i H_0 t} e^{-i H_1 t} \ket{I}.
\]
Clearly, if $\ket{I}$ is an eigenstate of both $H_0$ and $H_1$, then $r(t)$ is a phase and \eqref{eq:coherence_criterion} is satisfied, so the existence of a common eigenstate of these Hamiltonians (or, equivalently, of $\tilde{H}$ and $H_{\EE}$) is a sufficient condition. It is also a necessary condition, \emph{although the state $\ket{I}$ itself need not be a common eigenstate}. To understand this, write $H_0$ in terms of its spectral decomposition, with eigenvalues $\{\la_j^{(0)}\}$ and associated projection operators $\{\Pi_j^{(0)}\}$ (so that $H_0=\sum_j \la_j^{(0)} \Pi_j^{(0)}$), and similarly for $H_1$. Then
\beq
\label{rone}
r(t)=\sum_{j,k}e^{i(\la_j^{(0)}-\la_k^{(1)})t}\bra{I}\Pi_j^{(0)}\Pi_k^{(1)}\ket{I}.
\eeq
In general, $r(t)$ contains terms of different frequencies. However, in order for it to remain of unit magnitude for all $t$, only one frequency can appear; all others must be associated with vanishing coefficients. The easiest way for this to occur is if $\ket{I}$ is a common eigenstate of $H_0$ and $H_1$, in which case the double sum collapses to a single term. A more general possibility is if $\ket{I}$ is a linear combination of common eigenstates of $H_0$ and $H_1$ corresponding to the same energy difference. In terms of the original Hamiltonians $\tilde{H}$ and $H_{\EE}$, 
$\ket{I}$ must be a linear combination of eigenstates of $H_{\EE}$ and each of these must also be a degenerate eigenstate of $\tilde{H}$. Stated otherwise, we must be able to find a basis in which $\tilde{H}=C_{M} \oplus \tilde{H}'$ and $H_{\EE}=D_{M}\oplus H_{\EE}'$, 
where $C_{M}$ is an $M$-dimensional constant matrix (proportional to the identity)
and $D_{M}$ is an $M$-dimensional diagonal matrix; 
$\ket{I}$ can be any vector in the first $M$ dimensions.

Thus, preparing an initial state of the environment perfectly maintaining the coherence of the qubit at all times can only be done if both the interaction Hamiltonian and the environment self-evolution Hamiltonian exhibit a specific structure, namely they share an eigenstate. 

Can we characterize the rarity of such pairs among all pairs of Hamiltonians? From a mathematical point of view, it can be shown that pairs of Hermitian matrices sharing an eigenstate are a closed set with empty interior. 
Intuitively, this implies that pairs of Hermitian matrices with a common eigenstate are rare. This conclusion is also supported by the observation that even if both Hamiltonians share a common eigenstate, most perturbations will destroy this property. To see this, suppose that the Hamiltonians are written in a basis where $\tilde{H}$ is diagonal and suppose that only the first eigenstate is common to both Hamiltonians. Then the first row and column of $H_{\EE}$ are zero except for the $(1,1)$ element. In order for a perturbation of $H_{\EE}$ to preserve the common eigenstate, it (the perturbation) must have the same structure, so the real dimension of perturbations preserving this structure is less than the real dimension of all possible perturbations ($(N-1)^2+1$ vs $N^2$).

Notice that if the environment self-evolution Hamiltonian $H_{\EE}$ is zero, any eigenstate of the interaction Hamiltonian is a suitable initial state. Thus, it is the dynamics of the environment that usually prevents the existence of such an initial state. In the following section, we will make this situation explicit by computing the loss of coherence induced by adding dynamics to an otherwise-static environment.

{\em Environment evolution as a perturbation~~~}
In this section, we consider a solvable model due to Zurek \cite{Einselection} 
to which we add a new term $\id \otimes H_{\EE}$ to provide dynamics to an otherwise-static environment. 
We consider the case in which the environment is initially prepared in an eigenstate of $\tilde{H}$
and show that the self-evolution of the environment will destroy
the coherence of the system.
In this model, the environment consists of $n$ spin-1/2
particles. The dimension of the environment Hilbert space is $N=2^{n}$.
Every particle of the environment interacts with the system
through a $\sigma^{z}\sigma^{z}$ interaction. 
The eigenstates of the Pauli matrix $\sigma_{k}^{z}$ acting on the $k^{\rm{th}}$
spin are $\{\ket{0}_{k},\ket{1}_{k}\}$. The strength of the interaction
between the system and the $k^{\rm{th}}$ spin of the environment
is measured by a coupling constant $g_{k}\in\mathbb{R}$. Thus, the
Hamiltonian is
\beq
H  =  \sigma_{\SS}^{z}\otimes\sum_{k=1}^{n}g_{k}\sigma_{k}^{z}.
\label{e1}
\eeq

 The eigenstates of $\tilde{H}=\sum_{k=1}^{n}g_{k}\sigma_{k}^{z}$
are the states in which the $k^{\rm{th}}$ spin of the environment
is in an eigenstate of $\sigma_{k}^{z}$, i.e., the states $\ket{x}=\bigotimes_{k=1}^{n}\ket{x_{k}}_k$
where all $x_{k}\in\{0,1\}$. A convenient way to order these states
is to consider that $x$ is a number between $0$ and $N-1$ of which the
binary representation is $x={x_{1}x_{2}\dots x_{n}}$. Thus,
\[
\forall x\in\{0,1\}^{n}\quad\tilde{H}\ket{x}=\omega_{x}\ket{x}
\]
where $\om_x=\sum_{k=1}^n (-1)^{x_k} g_k$. Throughout this section, we will assume that the coupling constants $\{g_{k}\}$ are chosen so that the eigenvalues $\omega_{x}$ are distinct.

\begin{figure}[th]
 \centering
 \includegraphics[height=25mm,keepaspectratio=true]{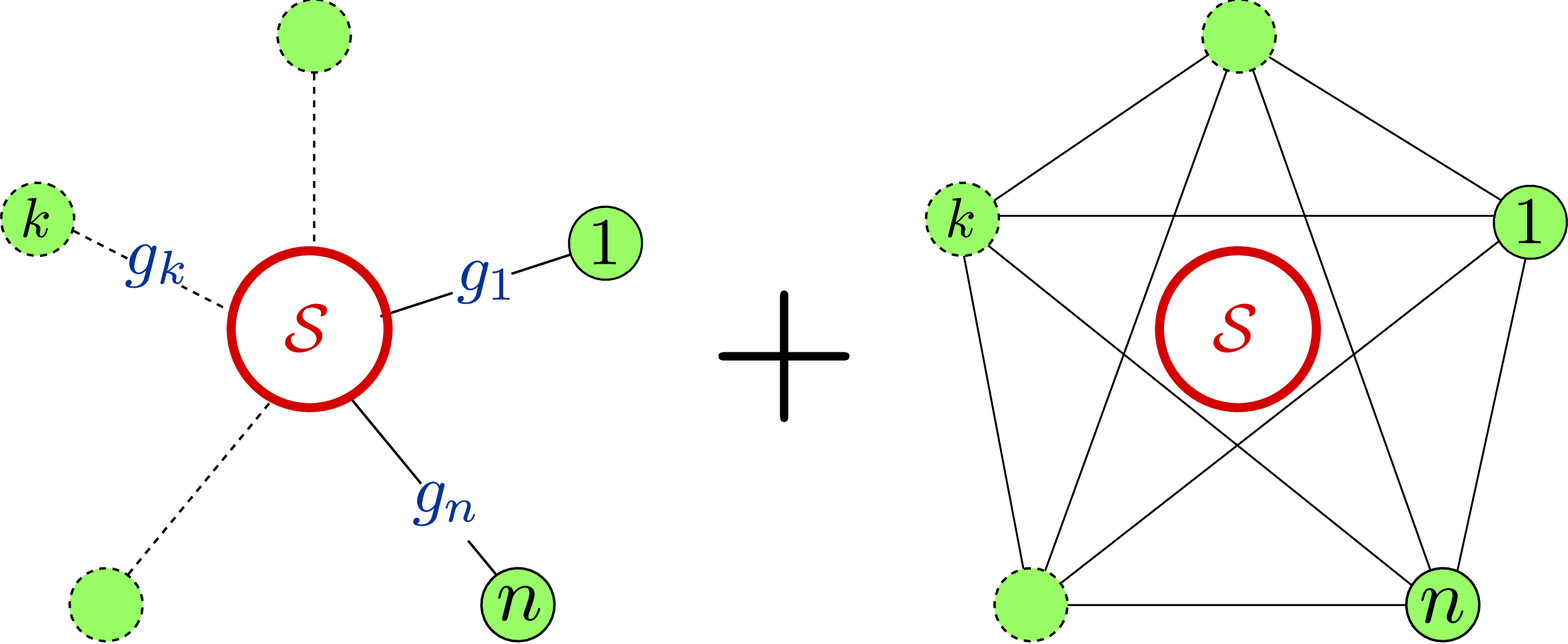}
 \caption{Decoherence model with dynamical environment}
 \label{Model}
\end{figure}

So far, the spins of the environment do not interact
with each other; thus, preparing the environment in any state $\ket{x}$, being an eigenstate of $\tilde{H}$, will give rise to coherent evolution of the system. We will now add a self-evolution of the environment
of the form
\[
H_{\EE}=\lambda\sum_{x,y\in\{0,1\}^{n}}\ket{x}\bra{y}
\]
where $\lambda\in\mathbb{R}$ is a perturbation parameter. This Hamiltonian
is proportional to the projector on $\ket{\Phi}=\textstyle{\frac{1}{\sqrt{N}}}\sum_{x\in\{0,1\}^{n}}\ket{x}=\bigotimes_{k=1}^{n}\ket{+_k}\bra{+_k}$, where $\ket{+_k}=\textstyle{\frac{1}{\sqrt{2}}}\ket{0_k}+\textstyle{\frac{1}{\sqrt{2}}}\ket{1_k}$:
\[
H_{\EE}=\lambda N\ket{\Phi}\bra{\Phi}\mbox{.}
\]
Clearly, the interaction and the environment Hamiltonians do not share
a common eigenstate because all eigenstates of $\tilde{H}$ have a
small overlap with $\ket{\Phi}$.

Suppose that the environment is prepared in the state
\begin{equation}
\ket{0}=\bigotimes_{k=1}^{n}\ket{0}_{k}\label{eq:ground_state}
\end{equation}
which is an eigenstate of $\tilde{H}$ with eigenvalue $\om_0$. 

Let us compute the decoherence
factor $r(t)$ for this initial state of the environment. Standard
perturbation theory for $\lambda\ll\min_{x,y\in\{0,1\}^{n}}|\omega_{x}-\omega_{y}|$
shows that
\[
|r(t)|^{2}=1-16\lambda^{2}\sum_{x\neq0}\frac{\sin^{4}\left(\frac{E_{x}-E_{0}}{2}t\right)}{\left(\omega_{0}-\omega_{x}\right)^{2}}.
\]
The time average of this quantity is given by
\[
\overline{|r(t)|^{2}}=1-6\lambda^{2}\sum_{x\neq0}\frac{1}{(\omega_{0}-\omega_{x})^{2}}\mbox{.} \]
Therefore, the time average is reduced by a factor proportional to
$\sum_{x\neq0}\frac{1}{(\omega_{0}-\omega_{x})^{2}}$, which characterizes
the density of energy levels near the unperturbed energy level.

{\em Imperfect preparation of the environment~~~}
Suppose now that the condition for coherent evolution in the original model \eqref{hamiltonian} is satisfied, so that there is at least one state for which the time evolution preserves the coherence of the system $\SS$. We have assumed implicitly our ability to prepare perfectly
the initial state of the environment. Notice that this strong assumption
is nonetheless weaker than requiring control over the environment dynamics
at all times. The interest of preparing the environment has already been studied 
from both a theoretical \cite{dajka-2009} and an experimental \cite{reilly2008ssq} point of view. However, preparing
the environment in a given initial state is a difficult task which
might only be achieved partially. Let us give a simple example in
the case of Zurek's model \eqref{e1}. When trying to prepare the state \eqref{eq:ground_state}, the goal is to prepare all spins
in the state $\alpha_{k}\ket{0}_{k}+\beta_{k}\ket{1}_{k}$ with
$\alpha_{k}=1$ and $\beta_{k}=0$. Suppose that we are only able
to ensure that $|\beta_{k}|^{2}\leq\varepsilon\ll1$ for all $k$, i.e.,
all spins of the environment are prepared with a small error. In that
case, the average value $\overline{|r(t)|^{2}}$ is bounded
by
\[
\overline{|r(t)|^{2}}\geq\left((1-\varepsilon)^{2}+\varepsilon^{2}\right)^{n}\stackrel{\varepsilon\ll1}{\longrightarrow}1-2n\varepsilon
\]
which is attained if $|\beta_{k}|^{2}=\varepsilon$ for all $k$. Thus,
for small independent errors on each of the $n$ spin of the environment,
the coherence loss is proportional to $n$. This particular example indicates
that even if an initial state allowing for decoherence-free evolution
exists, the difficulty to prepare it will grow with the size of the
environment, as one would expect intuitively.

{\em Imperfect control of the environment~~~}
We now address a situation which is in a sense opposite to the one just considered. Rather than having perfect control over the dynamics (so that a common eigenstate of $\tilde H$ and $H_{\EE}$ can be made to exist) but an imperfect ability to prepare the initial state, suppose that we can prepare perfectly any state we wish but that no such common eigenstate exists. Given that the coherence of $\SS$ cannot be preserved, is there an optimal choice of initial state, that is, one for which the ensuing decoherence is in some sense minimized? To address this question, we must first specify what we mean by optimal. Do we wish to minimize average decoherence, in which case we would not ``see" brief but significant drops in the coherence? Alternatively, do we wish to minimize the maximum decoherence, in which case a significant drop in coherence will make a state appear to be a bad choice, even though it may be good on average. In the following we will adopt this latter criterion.

\begin{figure}[ht]
 \centering
 \includegraphics[width=\columnwidth,keepaspectratio=true]{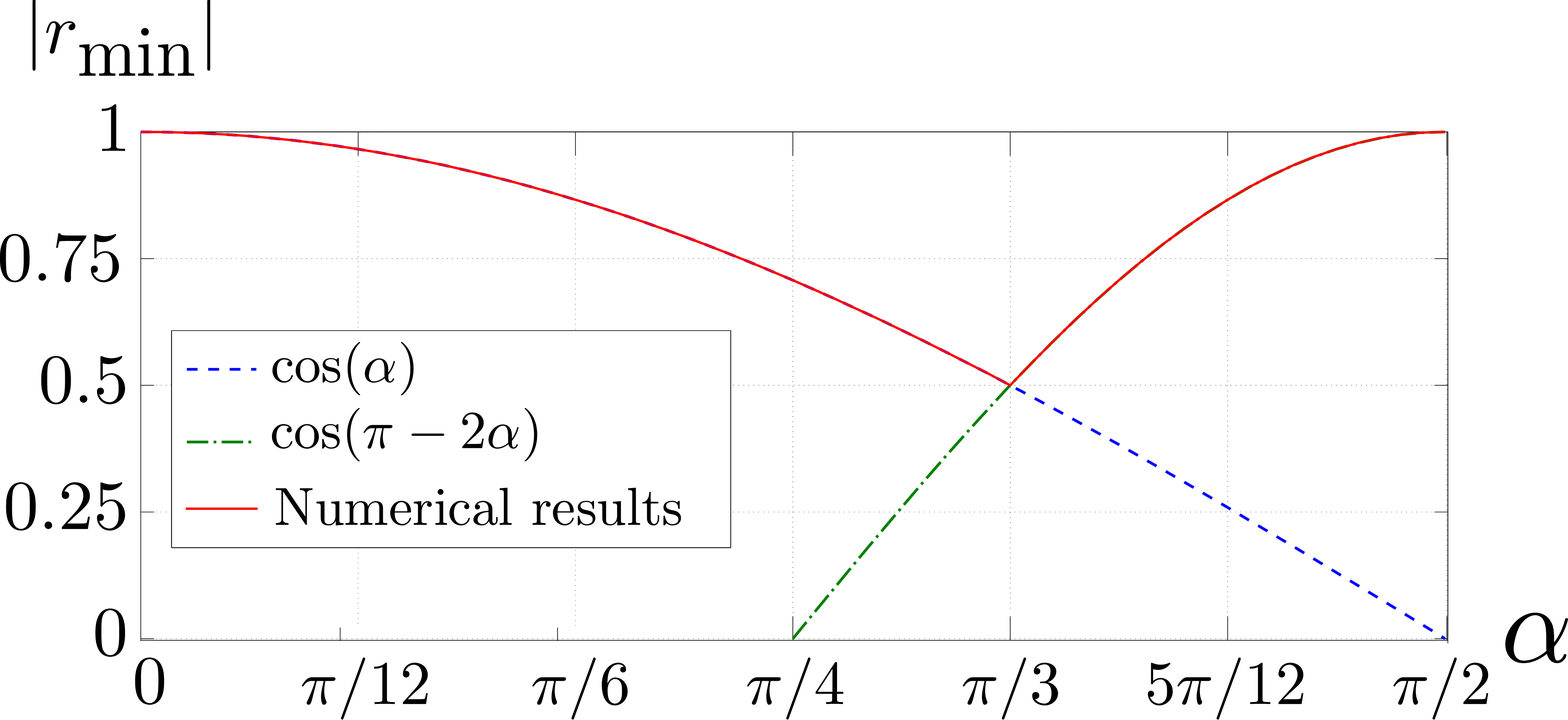}
 \caption{Numerical results}
 \label{NumResults}
\end{figure}

It is somewhat easier to use the combinations $H_{0,1}$ rather than $\tilde H$ and $H_{\EE}$. We wish to find the state $\ket{I}$ for which the minimum value of $|r(t)|$ is maximal, in the case where several frequencies are present in the sum \eqref{rone}. The general case appears difficult to analyze, but one might expect that the best choice of $\ket{I}$ is an eigenstate of one of the two Hamiltonians and a combination of two eigenstates of the other Hamiltonian. For instance, if $\Pi^{(0)}_1\ket{I}=(\Pi^{(1)}_1+\Pi^{(1)}_2)\ket{I}=\ket{I}$, then \eqref{rone} becomes
\[
r(t)=e^{i\la^1_1 t}\left( e^{-i\la^2_1 t}\bra{I}\Pi^{(1)}_1\ket{I}+e^{-i\la^2_2 t}\bra{I}\Pi^{(1)}_2\ket{I} \right)
\]
and
\[
|r_{\rm min}| = | \bra{I}\Pi^{(1)}_1\ket{I} - \bra{I}\Pi^{(1)}_2\ket{I} |.
\]
This expectation turns out  not to be the best choice, in general. To see this, we examine the simplest example in which no common eigenstate exists, namely, an environment consisting of a single qubit with $H_{0,1}$ describing its interaction with non-parallel magnetic fields of the same intensity. Let the precession frequency be $\om$ and let the direction of the magnetic fields corresponding to $H_0$ and $H_1$ be $\hat m_0 = (\sin \al,0,\cos\al)$ and $\hat m_1 = (-\sin \al,0,\cos\al)$, with $0<\al<\pi/2$, respectively. The initial state can be taken to be a spin aligned along a third direction, $\hat v = (\sin\theta\cos\phi,\sin\theta\sin\phi,\cos\theta)$, say. Its evolution according to each of the Hamiltonians is simple: the spin precesses around $\hat m_i$ with frequency $\om$ so $\ket{\varepsilon_i(t)}=\ket{\hat v_i(t)}$ where $\hat v_i(t)$ is $\hat v$ rotated about $\hat m_i$ by angle $\om t$. Then $r(t)=\vev{\varepsilon_0(t) | \varepsilon_1(t)}$ and $|r(t)| = \cos(\gamma(t)/2)$, where $\gamma(t)$ is the angle between $\hat v_0(t)$ and $\hat v_1(t)$. Thus we would like to find the vector $\hat v$ for which the maximum angle between $\hat v_0(t)$ and $\hat v_1(t)$ as they precess is minimized. If, as was conjectured above, we choose $\hat v=\hat m_0$, then it is easy to see that the maximum angle is the lesser of $4\al$ and $2\pi-4\al$, reached after half a precession. Although it is surprisingly difficult to find this maximum angle for an arbitrary $\hat v$, the optimal choice turns out to depend on the angle $\al$ as indicated in figure \ref{NumResults}. 

If $\al \leq \pi/3$, it is best to choose $\hat v=\hat y$. perpendicular to $\hat m_0$ and $\hat m_1$, resulting in $\gamma_{\rm max}=2\al$ (attained after a quarter-rotation) and
\[
|r_{\rm min}| =\cos \al.
\]

If $\al \geq \pi/3$, an optimal choice is $\hat v$ collinear with $\hat m_0$ (or $\hat m_1$), resulting in $\gamma_{\rm max}=2\pi-4\al$ (attained after a half-rotation) and
\[
|r_{\rm min}| =\cos (\pi-2\al).
\]

Thus, the optimal value of $|r_{\rm min}|$ is close to unity if the directions of the two magnetic fields are almost parallel. More surprising is the fact that the best choice of initial state is far from what one would naively have guessed. In the case in which $\al \leq \pi/3$, the optimal choice is $\hat v = \hat y$, which does not correspond to an eigenstate of either Hamiltonian.


To sum up, two regimes emerge. On the one hand, for $\alpha > \pi/3$, the system-environment interaction prevails whereas the self-evolution of the environment is only a perturbation. In this regime, the system and the environment play symmetric roles. Reducing decoherence in this case boils down to minimizing the entanglement. The optimal initial states are thus the pointer states of the environment, i.e. the eigenstates of $\tilde{H}$. On the other hand, for $\alpha < \pi/3$, the self-evolution Hamiltonian $H_\EE$ dominates over the interaction Hamiltonian $\tilde{H}$. However, the evolution \emph{of the quantum system} relies essentially on the interaction. Thus, for the evolution of the system, $\tilde{H}$, no matter how small, \emph{cannot} be considered a perturbation with respect to $H_{\EE}$. In that case, the evolution of the environment is dominated by its own dynamics but the impact on the quantum system is mediated by the interaction Hamiltonian. Hence, finding an analytical criterion characterizing an initial state of the environment that optimally limits the subsequent decoherence of the system remains an unresolved challenge. 

\acknowledgments{We thank the National Science and Engineering Research Council of Canada for funding, and David Poulin for useful discussions.}




\bibliography{biblio}

\begin{thebibliography}{8}
\expandafter\ifx\csname natexlab\endcsname\relax\def\natexlab#1{#1}\fi
\expandafter\ifx\csname bibnamefont\endcsname\relax
  \def\bibnamefont#1{#1}\fi
\expandafter\ifx\csname bibfnamefont\endcsname\relax
  \def\bibfnamefont#1{#1}\fi
\expandafter\ifx\csname citenamefont\endcsname\relax
  \def\citenamefont#1{#1}\fi
\expandafter\ifx\csname url\endcsname\relax
  \def\url#1{\texttt{#1}}\fi
\expandafter\ifx\csname urlprefix\endcsname\relax\def\urlprefix{URL }\fi
\providecommand{\bibinfo}[2]{#2}
\providecommand{\eprint}[2][]{\url{#2}}

\bibitem[{\citenamefont{$\dot{\mbox{Z}}$urek}(1982)}]{Einselection}
\bibinfo{author}{\bibfnamefont{W.~H.} \bibnamefont{$\dot{\mbox{Z}}$urek}},
  \bibinfo{journal}{Phys. Rev. D} \textbf{\bibinfo{volume}{26}},
  \bibinfo{pages}{1862} (\bibinfo{year}{1982}).

\bibitem[{\citenamefont{Schlosshauer}(2007)}]{Schlosshauer2007}
\bibinfo{author}{\bibfnamefont{M.}~\bibnamefont{Schlosshauer}},
  \emph{\bibinfo{title}{Decoherence and the Quantum to Classical Transition}}
  (\bibinfo{publisher}{Springer-Verlag Berlin}, \bibinfo{year}{2007}).

\bibitem[{\citenamefont{Paz and
  $\dot{\mbox{Z}}$urek}(2001)}]{LesHouchesPazZurek}
\bibinfo{author}{\bibfnamefont{J.~P.} \bibnamefont{Paz}} \bibnamefont{and}
  \bibinfo{author}{\bibfnamefont{W.~H.} \bibnamefont{$\dot{\mbox{Z}}$urek}},
  \emph{\bibinfo{title}{Coherent Matter Waves, Lectures from the 72nd Les
  Houches Summer School, 1999}} (\bibinfo{publisher}{Springer-Verlag, Berlin},
  \bibinfo{year}{2001}), pp. \bibinfo{pages}{533--614}.

\bibitem[{\citenamefont{Nielsen and Chuang}(2000)}]{NielsenChuang}
\bibinfo{author}{\bibfnamefont{M.~A.} \bibnamefont{Nielsen}} \bibnamefont{and}
  \bibinfo{author}{\bibfnamefont{I.~L.} \bibnamefont{Chuang}},
  \emph{\bibinfo{title}{Quantum Information and Quantum Information}}
  (\bibinfo{publisher}{Cambridge University Press}, \bibinfo{year}{2000}).

\bibitem[{\citenamefont{Hornberger}(2009)}]{Hornberger2009}
\bibinfo{author}{\bibfnamefont{K.}~\bibnamefont{Hornberger}},
  \emph{\bibinfo{title}{Entanglement and Decoherence}}
  (\bibinfo{publisher}{Springer Berlin / Heidelberg}, \bibinfo{year}{2009}),
  vol. \bibinfo{volume}{768} of \emph{\bibinfo{series}{Lecture Notes in
  Physics}}, pp. \bibinfo{pages}{221--276}.

\bibitem[{\citenamefont{Cucchietti et~al.}(2005)\citenamefont{Cucchietti, Paz,
  and Zurek}}]{Cucchietti2005}
\bibinfo{author}{\bibfnamefont{F.~M.} \bibnamefont{Cucchietti}},
  \bibinfo{author}{\bibfnamefont{J.~P.} \bibnamefont{Paz}}, \bibnamefont{and}
  \bibinfo{author}{\bibfnamefont{W.~H.} \bibnamefont{Zurek}},
  \bibinfo{journal}{Phys. Rev. A} \textbf{\bibinfo{volume}{72}},
  \bibinfo{pages}{052113} (\bibinfo{year}{2005}).

\bibitem[{\citenamefont{Dajka et~al.}(2009)\citenamefont{Dajka, Mierzejewski,
  Luczka, and Hanggi}}]{dajka-2009}
\bibinfo{author}{\bibfnamefont{J.}~\bibnamefont{Dajka}},
  \bibinfo{author}{\bibfnamefont{M.}~\bibnamefont{Mierzejewski}},
  \bibinfo{author}{\bibfnamefont{J.}~\bibnamefont{Luczka}}, \bibnamefont{and}
  \bibinfo{author}{\bibfnamefont{P.}~\bibnamefont{Hanggi}},
  \emph{\bibinfo{title}{Dephasing of qubits by the schr\"odinger cat}}
  (\bibinfo{year}{2009}), \urlprefix\url{arXiv.org:0905.2569}.

\bibitem[{\citenamefont{Reilly et~al.}(2008)\citenamefont{Reilly, Taylor,
  Petta, Marcus, Hanson, and Gossard}}]{reilly2008ssq}
\bibinfo{author}{\bibfnamefont{D.}~\bibnamefont{Reilly}},
  \bibinfo{author}{\bibfnamefont{J.}~\bibnamefont{Taylor}},
  \bibinfo{author}{\bibfnamefont{J.}~\bibnamefont{Petta}},
  \bibinfo{author}{\bibfnamefont{C.}~\bibnamefont{Marcus}},
  \bibinfo{author}{\bibfnamefont{M.}~\bibnamefont{Hanson}}, \bibnamefont{and}
  \bibinfo{author}{\bibfnamefont{A.}~\bibnamefont{Gossard}},
  \bibinfo{journal}{Science} \textbf{\bibinfo{volume}{321}},
  \bibinfo{pages}{817} (\bibinfo{year}{2008}).

\end{thebibliography}

\end{document}